\documentstyle[graphicx]{mn}

\title{An XMM-Newton observation of the extreme Narrow Line Seyfert~1 
Galaxy, Mrk~359}
\author[P.T. O'Brien et al.]
       {P.T. O'Brien,$^1$ K. Page,$^1$ J.N. Reeves,$^1$ K.
Pounds,$^1$ M.J.L. Turner,$^1$ 
\newauthor 
E.M. Puchnarewicz$^2$\\
$^1$X-ray Astronomy Group, Department of Physics and Astronomy,
University of Leicester, LE1 7RH, U.K.\\
$^2$Mullard Space Science Laboratory, University College London,
Holmbury St. Mary, Dorking, Surrey, RH5 6NT, UK}

\date{2001 June}

\begin{document}

\maketitle

\label{firstpage}

\begin{abstract}
We present {\it XMM-Newton} observations of Mrk~359, the first Narrow
Line Seyfert 1 galaxy discovered. Even among NLS1s, Mrk~359 is an
extreme case with extraordinarily narrow optical emission lines. The
{\it XMM-Newton} data show that Mrk~359 has a significant soft X-ray
excess which displays only weak absorption and emission features. The
(2--10)~keV continuum, including reflection, is flatter than the
typical NLS1, with $\Gamma \approx 1.84$. A strong emission line of 
equivalent width $\approx 200$~eV is also observed, centred near 6.4
keV. We fit this emission with two line components of approximately
equal strength: a broad iron-line from an accretion disc and a narrow,
unresolved core. The unresolved line core has an equivalent width of
$\approx 120$~eV and is consistent with fluorescence from neutral
iron in distant reprocessing gas, possibly in the form of a `molecular
torus'. Comparison of the narrow-line strengths in Mrk~359 and other
low--moderate luminosity Seyfert~1 galaxies with those in QSOs
suggests that the solid angle subtended by the distant reprocessing
gas decreases with increasing AGN luminosity.

\end{abstract}


\section{Introduction}

Exploring the extremes of parameter space is an invaluable way of
probing the physical processes occurring in any class of astronomical
object. We adopt this approach here and discuss the {\it XMM-Newton}
X-ray spectra of the nearby Seyfert 1 galaxy Mrk~359 ($z = 0.0174$).
Davidson \& Kinman (1978) noted that lines from both the Narrow Line
Region (NLR) and the Broad Line Region (BLR) are unusually narrow in
Mrk~359. They also noted the presence of strong high-ionisation lines
and suggested Mrk~359 merited further study. The narrowness of the
emission lines combined with the presence of strong high-ionisation
lines led Osterbrock \& Dahari (1983) to classify Mrk~359 as a Narrow
Line Seyfert 1 galaxy (NLS1), the first such object discovered.
Veilleux (1991) found that the FWHM of the forbidden lines (NLR) is
$\approx 135$ km s$^{-1}$ while the permitted lines (BLR) have FWHM
$\approx 800$ km s$^{-1}$. Even for NLS1s, these are extraordinarily
narrow line widths. Both the NLR and BLR profiles are also unusual
among the sample studied by Veilleux in that they show no
substructure.

NLS1s have become a prime target for X-ray observation due to their
extreme variability characteristics and unusual spectral shape (e.g.
Boller, Brandt \& Fink 1996). Mrk~359 has relatively little previous
X-ray data. It was first detected during the {\it Einstein} slew
survey (Elvis et al. 1992). Assuming it has a `normal' 2--10~keV
spectral index, Walter \& Fink (1993) suggest Mrk~359 has a moderately
strong soft X-ray excess based on the {\it ROSAT} all sky survey data.
Boller et al. (1996) found it had a fairly flat continuum for the NLS1
class in the {\it ROSAT} bandpass ($\Gamma = 2.4 \pm 0.1$).

In this paper we present {\it XMM-Newton} observations of Mrk~359. Due
to the combination of the broad bandpass, spectral resolution and
high throughput of the {\it XMM-Newton} instrumentation, these data
are a considerable improvement on previous observations. We find
strong line emission, centred at 6.4~keV in the rest-frame of Mrk~359,
superimposed on a type-1 AGN continuum consisting of a
high-energy powerlaw and a strong soft X-ray excess.

\section[]{XMM-Newton Observations}

Although actually a guaranteed time target, Mrk~359 was observed in
revolution 107 (9 July 2000) during the early calibration and
performance verification phase of the mission. The European Photon
Imaging Camera (EPIC) PN (Str\"uder et al. 2001) and EPIC MOS (Turner
et al. 2001) exposure times were 7.3 and 4~kecs respectively. The
Reflection Grating Spectrograph (RGS; den Herder et al. 2001) exposure
time was 27~ksecs.

The EPIC MOS and PN cameras were operated in full-frame observing mode
using the medium filters. Event lists output from the standard {\sc
emchain} and {\sc epchain} scripts were further filtered using the
{\sc SAS} (Science Analysis Software) {\sc xmmselect} task. Only X-ray
events corresponding to patterns 0--12 (similar to {\it ASCA} event
grades 0--4) were selected for the MOS exposures. For the PN patterns
0--4 (singles and doubles) were used. Hot pixels and electronic noise
were rejected during data processing and the low-energy cutoff for
spectral extraction was set at 200~eV.

Source spectra were extracted from the EPIC images using a circular
source region of diameter 1 arcminute centred on the observed source
position. Background spectra were derived from adjacent `blank sky'
regions. Before further analysis the EPIC spectra were binned to give
a minimum of 20 counts per bin. The {\sc Xspec v11.0} software package
was used to calibrate the background-subtracted EPIC spectra using the
most recent camera response matrices derived from ground-based and
in-orbit data. Errors on fitted parameters are quoted at the 90\%
confidence level.

The RGS data were processed using the standard SAS {\sc rgsproc}
script and calibrated using response matrices derived using the {\sc
rgsrmfgen} task.

% Table 1
\begin{table}
 \centering
% \begin{minipage}{110mm}
  \caption{Fits to {\it XMM-Newton} 2--10~keV data for Mrk~359.
$^a$Excluding 5.5--7~keV region.
$^b$Rest-frame energy of emission line
(intrinsic line-width, $\sigma < 0.08$ keV).
}
\begin{tabular}{@{}lllccr@{}}
Camera & Fit & Model & $\Gamma$ & E$^b$ (keV) & 
$\chi^2$/dof \\
\\
MOS & 1$^a$ & PL & $1.77 \pm 0.09$ & -- & 56/69 \\
PN  & 1$^a$ & PL & $1.91 \pm 0.05$ & -- & 164/167 \\
MOS+PN & 1$^a$ & PL & $1.88 \pm 0.04$ & -- & 223/237 \\
PN & 2 & PL+GA & $1.88 \pm 0.04$ & $6.41 \pm 0.05$
& 190/192 \\
MOS+PN & 2 & PL+GA & $1.85 \pm 0.04$ & $6.43 \pm 0.03$
& 259/270 \\
\end{tabular}
%\end{minipage}
\end{table}

\section{Spectral Analysis}

As usual with AGN, the first model fitted to Mrk~359 was that of a
single absorbed powerlaw. The absorbing column density was fixed at the
Galactic value of $N_h = 4.79 \times 10^{20}$ cm$^{-2}$. Fitting to the
joint EPIC MOS+PN data over the entire 0.2--10~keV bandpass provides
a poor fit ($\chi^2_{\nu} = 1097/743$). To examine the detailed
spectral shape, the fitting procedure was therefore split in two,
first fitting over 2--10~ keV and then over the broad-band 0.2--10~keV.

\subsection{The 2--10 keV spectrum}

\begin{figure}
\centering
\includegraphics[width=5.5cm,angle=-90]{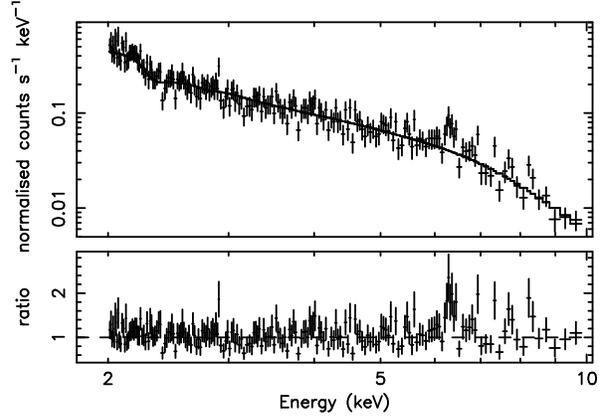}
\caption{The 2--10 keV EPIC PN spectrum of Mrk~359 fitted with a
powerlaw. The powerlaw fit excludes the region from 5.5--7 keV 
(fit 1 in Table 1). The lower plot shows the ratio of the fitted
powerlaw to the data.}
\label{hardpow}
\end{figure}

A powerlaw was fitted over 2--10~keV excluding the iron line region
(5.5--7 keV). This fit, called fit 1 in Table 1, provides an
acceptable fit to either the MOS or PN data separately or when
combined (e.g., $\chi^2_{\nu} = 0.93$ for MOS+PN) giving a photon
index $\Gamma \approx 1.9$. Fit 1 for the PN is shown in Fig.\
\ref{hardpow}.

% Fig. 1

% Fig. 2
\begin{figure}
\centering
\includegraphics[width=5.5cm,angle=-90]{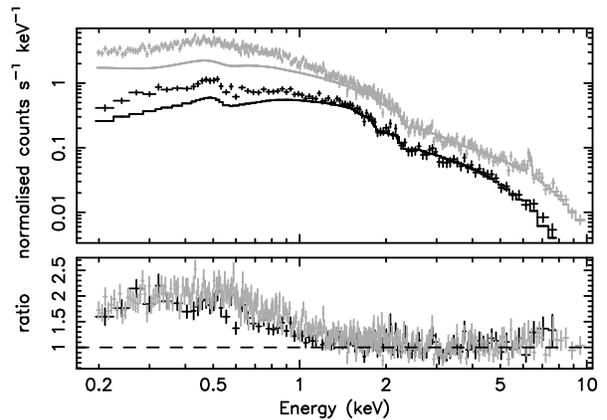}
\caption{The 0.2--10 keV EPIC MOS and PN spectra of Mrk~359. The extrapolated
2--10 keV powerlaw+Gaussian line model (fit 2 in Table 1) is also
shown. A strong soft excess is present below $\approx 2$ keV.}
\label{broadband}
\end{figure}

A clear `feature' is seen near 6.4 keV in the PN data (Fig.\
\ref{hardpow}), so a Gaussian emission line was added (fit 2 in Table
1). This significantly improves the fit compared to that of a
powerlaw fitted over the entire 2--10 keV band ($\Delta
\chi^2 = 23$ for 3 additional parameters). The emission line is
not well defined in the MOS data, but using both datasets gives a fit
statistically consistent to that of the PN alone. Adding the emission
line flattens the PN hard continuum slope slightly to $\Gamma = 1.88$.
The rest-frame energy for the emission line in the PN spectrum is $6.41
\pm 0.05$ keV, consistent with that expected for fluorescence from
neutral iron. The fitted Gaussian line is unresolved ($\sigma <80$ eV)
and is very strong, with an EW of $216 \pm 73$~eV in the PN spectrum.
Adding a second Gaussian line does not improve the fit, but there are
indications of a broader line component in the data residuals (Fig.\
\ref{hardpow}). We discuss more detailed iron line models below.

\subsection{The 0.2--10 keV spectrum}
\begin{figure}
\centering
\includegraphics[width=5.5cm,angle=-90]{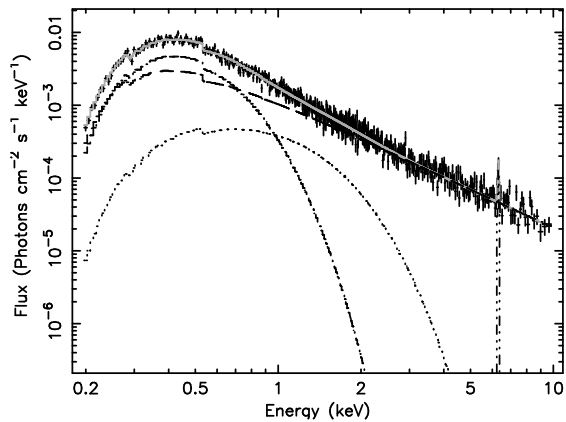}
\caption{The unfolded 0.2--10 keV EPIC PN spectrum of Mrk~359 fitted using
a powerlaw, two blackbodies and a Gaussian emission line (fit 4 in
Table 2).}
\label{finalconfit}
\end{figure}

Extrapolating the best-fit 2--10 keV powerlaw (fit 2 in Table 1) back
to 0.2~keV clearly reveals a broad ``soft excess'' in both the PN and
MOS spectra (Fig.\ \ref{broadband}). To fit the broad-band spectrum,
we fixed the iron-line parameters and used blackbody (BB) components to
parameterise the soft excess. Using a single blackbody provides a
reasonable fit to the soft emission (fit 3 in Table 2) but the
powerlaw slope steepens. Using two blackbodies significantly improves
the fit ($\Delta \chi^2 = 20$ for 2 additional parameters; fit 4 in
Table 2) and hardens the fitted powerlaw. The unfolded spectrum
corresponding to fit 4 is shown in Fig.\ \ref{finalconfit}.

% Fig. 3

% Fig. 4
\begin{figure}
\centering
\includegraphics[width=5.5cm,angle=-90]{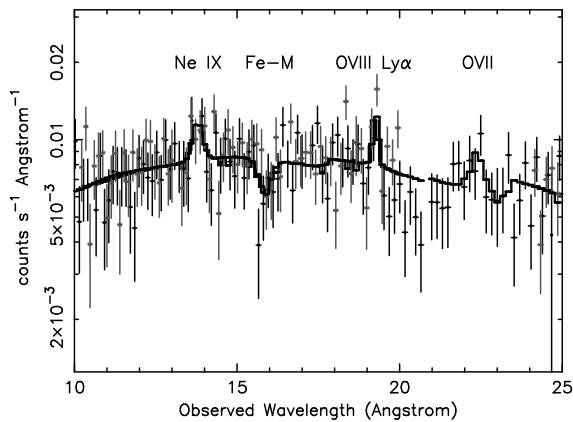}
\caption{The RGS spectrum of Mrk~359 showing the significant spectral
features. See text for details.}
\label{rgsfig}
\end{figure}

% Table 2
\begin{table*}
 \centering
 \begin{minipage}{170mm}
  \caption{Fits to {\it XMM-Newton} 0.2--10~keV data for Mrk~359.
$^a$Blackbody temperature (kT).
$^b$Rest-frame energy of emission line.
$^c$Intrinsic line-width (1-sigma).
$^d$Fit to ionised reflection model (Nayakshin et al. 2000) -- see
text for details.
$^f$Frozen.
}
\begin{tabular}{@{}lllccccr@{}}
Camera & Fit & Model & $\Gamma$ & BB$^a_1$ (keV) & BB$^a_2$ (keV) & 
E$^b$/$\sigma$$^c$ (keV) & 
$\chi^2$/dof \\
\\
MOS+PN & 3 & PL+BB+GA & $1.93 \pm 0.02$ & $0.131\pm0.002$ & -- &
$6.41^{f}$ / $0.01^{f}$ & 775/742\\
MOS+PN & 4 & PL+2$\times$BB+GA & $1.74 \pm 0.06$ & $0.120\pm0.003$ &
$0.310\pm0.026$ & $6.41^{f}$ / $0.01^{f}$ & 755/740\\
PN & 5 & Xion$^d$+BB+GA & $1.84\pm 0.02$ & $0.134\pm 0.003$ & -- &
$6.41^{f}$ / $0.01^{f}$ & 565/528 \\
\end{tabular}
\end{minipage}
\end{table*}

Using fit 4 in Table 2, the 0.2--10 keV flux 
of Mrk~359 is $1.26 \times 10^{-11}$ erg cm$^{-2}$ s$^{-1}$,
corresponding to a luminosity of $1.7
\times 10^{43}$ erg s$^{-1}$ (assuming $H_0 = 50$ km s$^{-1}$
Mpc$^{-1}$ and $q_0 = 0$). Most of this luminosity, $1.0 \times
10^{43}$ erg s$^{-1}$, is in the 0.2--2 keV band, distributed equally
between the powerlaw and multiple blackbody components. The 2--10 keV
luminosity of $7.0 \times 10^{42}$ erg s$^{-1}$ is dominated by the
powerlaw component in the fit.

The soft X-ray luminosity of Mrk~359 during the {\it XMM-Newton}
observation was more than a factor of two lower than the luminosity
derived by Boller et al. (1996) from the 15 July 1992 {\it ROSAT}
data. They note, however, that Mrk~359 varied by a factor of 1.5 in
just 10 hours. Fitting a powerlaw to the {\it XMM-Newton} data over
the 0.2--2.4 keV band gives $\Gamma = 2.34 \pm 0.02$, identical to the
value of $2.4 \pm 0.1$ derived from the {\it ROSAT} data. Thus,
although somewhat fainter, we have no evidence for spectral shape
changes in Mrk~359 over a timescale of eight years.

No large amplitude continuum variability was seen during the RGS
observation, but the continuum did vary by $\pm 15$\% on timescales of
several thousand seconds. This behaviour is consistent with that of
other NLS1s (e.g., Boller et al. 1996).

The RGS spectrum is of relatively low S/N, but the overall shape is
consistent with that of the EPIC spectra. No significant neutral
intrinsic absorption is detected (intrinsic $N_h < 1.24
\times 10^{20}$ cm$^{-2}$) nor evidence for a `conventional warm
absorber', with limits on the \hbox{O\,{\sc vii}} and \hbox{O\,{\sc
viii}} edge optical depths of $\tau < 0.1$.

A weak absorption trough is detected centred at $15.6 \pm 0.1$~\AA\
(rest-frame) with an EW $= 6.5\pm 3.0$ eV (Fig.\ \ref{rgsfig}),
similar in location and strength to those seen in IRAS~13349+2438
(Sako et al. 2001) and Mrk~509 (Pounds et al. 2001), which were
attributed to absorption in iron M ions. Narrow emission features are
detected at $13.55\pm 0.07$~\AA\ and $21.98\pm
0.13$~\AA\ (rest-frame) corresponding to the \hbox{Ne\,{\sc IX}} and
\hbox{O\,{\sc VII}} triplets.
The S/N is too low to resolve these features but their total EWs are $
12\pm 6$~eV and $6\pm 4$~eV respectively. The
\hbox{O\,{\sc viii}} Ly$\alpha$ line is also seen
at $18.94\pm 0.05$~\AA\ (rest-frame), with an EW $= 5.0\pm 2.5$~eV.
Any associated \hbox{O\,{\sc viii}} radiative recombination continuum
is less than half the strength of the Ly$\alpha$ emission line. This
ratio suggests that, as in NGC~4151 (Ogle et al. 2001), a significant
fraction of the \hbox{O\,{\sc viii}} emission may come from a hot,
collisional ionised rather than photoionised plasma, possibly
associated with a hot medium pressure-confining the NLR clouds.

% Fig. 5a,b
\begin{figure*}
\centering
\includegraphics[width=5.5cm,angle=-90]{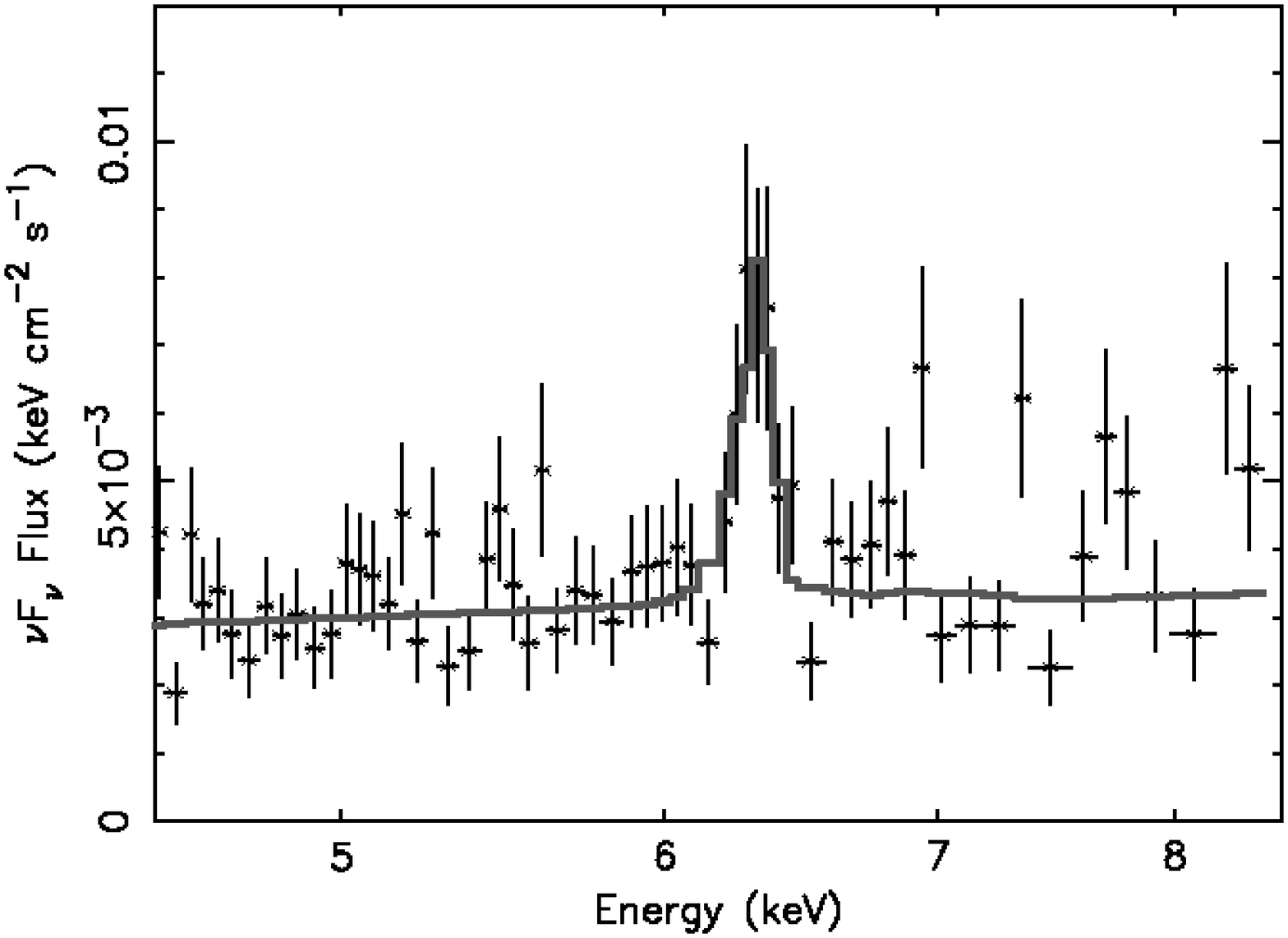}\hspace*{1cm}
\includegraphics[width=5.5cm,angle=-90]{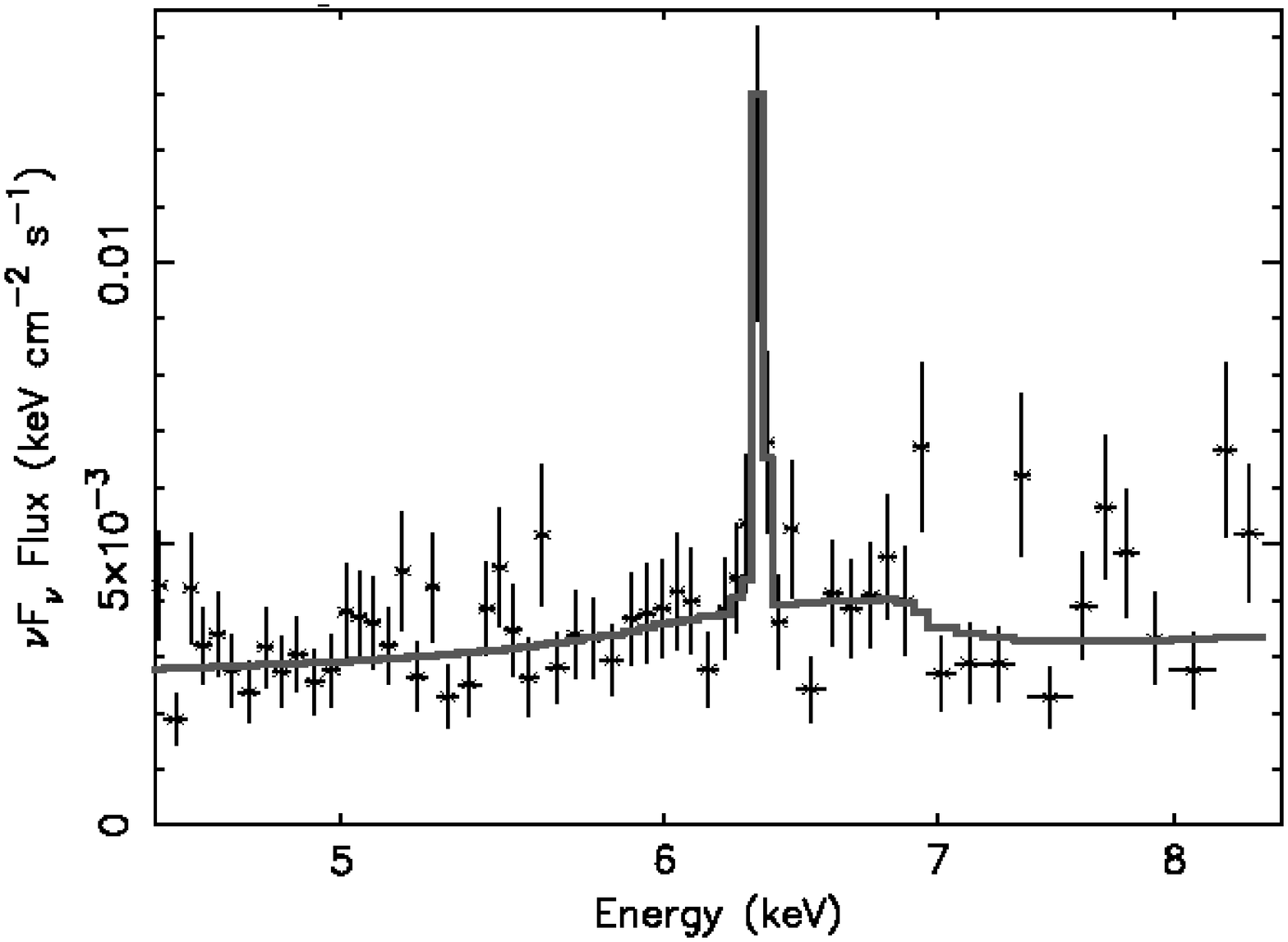}
\caption{Accretion disc model (Nayakshin et al. 2000) fits
to the 2--10 keV EPIC PN spectrum of Mrk~359. (a) face-on truncated
disc with inner radius 50~R$_s$ and accretion rate 1\% Eddington plus a
narrow 6.4~keV Gaussian core ($\sigma =0.01$~keV); (b) 30 degree
inclination disc extending down to 3~R$_s$ with accretion rate 30\%
Eddington plus a narrow 6.4~keV Gaussian core ($\sigma =0.01$).}
\label{discfits}
\end{figure*}

\section{The 6.4 keV iron emission line}

The 6.4~keV emission line in Mrk~359 appears much stronger (EW
$\approx 200$~eV) than the typical EW $50$--100~eV for `narrow' iron
lines seen in other Seyfert~1 galaxies (e.g. Reeves et al. 2001;
Yaqoob et al. 2001; Kaspi et al. 2001; Pounds et al. 2001).
Attributing all of the line emission in Mrk~359 to fluoresence from
cool reprocessing gas (e.g., a torus) illuminated by the central hard
X-ray continuum would require that the gas subtends an unusually large
solid angle (e.g. Ghisellini, Haardt \& Matt 1994), for which we have
no other evidence. We therefore investigated whether iron emission
from an accretion disc could be contributing to the observed line, as
expected in Seyfert galaxies (see Fabian et al. 2000 and references
therein). 

To quantify the disc contribution, we compared the Mrk~359 EPIC PN
2--10~keV data with an accretion disc model for a non-rotating black
hole. The generic `Xion' model used here is described in detail in
Nayakshin, Kazanas \& Kallman (2000). The `lamppost' geometry is
assumed for the illuminating hard X-ray source. The X-ray to disc
luminosity ratio was fixed at 0.3, which is reasonable for the
spectral energy distribution of Mrk~359. Two basic accretion disc
models were considered: a `truncated accretion disc' which produces an
intrinsically narrow emission line and a `normal accretion disc'
producing a broad line.

To explain the observed emission line as an accretion disc line which
is masquerading as a narrow line we must somehow remove the broad
wings commonly associated with such a line. The simplest way to do
this is to truncate the disc, removing the inner higher-velocity gas.
Our truncated disc model is shown in Fig.\ \ref{discfits}(a). For this
model $\chi^2_{\nu} = 189/193$. The inner and outer radii of the disc
are 50 and 1000 R$_s$ (Schwarzschild radii) respectively, and the
height of the X-ray source above the disc surface is 30 R$_s$. The
accretion rate is 0.01 Eddington. To keep the line centred at 6.4~keV,
the accretion rate must be low for this model ($< 0.05$) in order to
keep iron predominantly neutral. To explain the observed line strength
requires a super-solar iron abundance, which we fixed at 5 times
solar. To fit the peak of the line, the model shown in Fig.\
\ref{discfits}(a) includes an additional, narrow, unresolved Gaussian
line ($\sigma =0.01$~keV). This narrow-line component has an EW of
62~eV.

Although the truncated disc model provides a statistically acceptable
fit, it does not appear physically plausible. Truncated disc models
are more commonly proposed to explain soft X-ray transient (SXT)
systems, a sub-class of Low mass X-ray binaries (e.g., Dubus, Hameury
\& Lasota 2001). These models have some similarities to those for AGN,
including the possible influence of X-ray irradiation of the disc
surface. However, in SXTs the disc is truncated when the system is in
the `low state', which may be inappropriate for an NLS1. As noted
above, a low accretion rate is required for this model ($< 0.05$) to
keep the iron neutral. To explain the bolometric luminosity ($\approx
10^{44}$ erg s$^{-1}$) such a low accretion rate implies a 
black hole mass $\ga 10^8$ $M_{\sun}$. This would be a relatively
high mass for a low luminosity AGN like Mrk~359 (c.f., Ferrarese et
al. 2001), and is also contrary to the idea that NLS1s are relatively
low-mass systems running at high accretion rates (e.g., Pounds, Done
\& Osborne 1995). The generic X-ray properties of NLS1s, and in
particular their variability (Boller et al. 1996), strongly suggests
that they have discs extending down to the minimum allowed radius
(i.e., 3~R$_s$ for a non-rotating black hole).

To avoid these difficulties, an accretion disc model in which the disc
extends down to 3~R$_s$ was calculated and is shown in Fig.\
\ref{discfits}(b). For this model $\chi^2_{\nu} = 187/193$. The
parameters of the model are not well constrained, so we fixed the disc
inclination angle at 30 degrees, the height of the X-ray source above
the disc surface at 10 R$_s$ and the accretion rate at 30\% Eddington.
Such an accretion rate, which we consider reasonable for an NLS1,
results in an ionised disc although we cannot rule out an
essentially neutral broad line. A separate narrow Gaussian component
($\sigma = 0.01$~keV) is definitely required to fit the line core, and
has an EW of 127~eV. Thus, allowing for a contribution from a broad
iron line results in a narrow-line EW similar to that for the
narrow-lines in Seyfert 1 galaxies of comparable X-ray luminosity,
such as NGC~3783 (Kaspi et al. 2001) and NGC~5548 (Yaqoob et al.
2001).

% Fig. 5
\begin{figure}
\centering
\includegraphics[width=5.5cm,angle=-90]{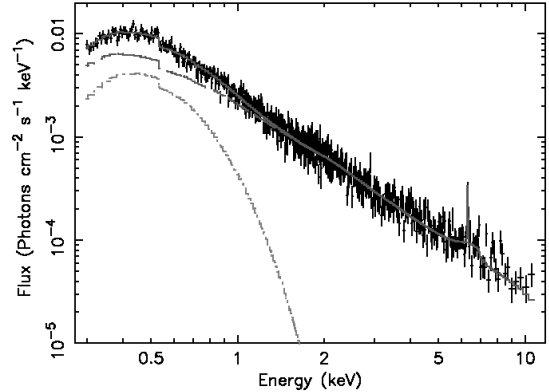}
\caption{The unfolded 0.2--10 keV EPIC PN spectrum of Mrk~359 fitted using
a single blackbody and the ionised accretion disc plus narrow Gaussian
emission-line model shown in Fig.\ \ref{discfits}(b) -- fit 5 in Table
2.}
\label{discreflect}
\end{figure}

Ionised accretion discs, such as that shown in Fig.\ \ref{discfits}(b),
will not only contribute to the iron line but also to the soft X-ray
emission due to reflection. This is illustrated in Fig.\
\ref{discreflect} where we extrapolate the disc model from Fig.\
\ref{discfits}(b) down to 0.2~keV. Adding a single blackbody to this
model provides a good fit to the broadband spectrum (fit 5 in Table 2),
removes the need for the hotter blackbody component used in fit 4
in Table 2 and recovers the powerlaw index of the 2--10~keV fit 2 in
Table 1.

\section{Conclusions}

The {\it XMM-Newton} spectrum of Mrk~359 reveals a strong iron
emission line centred at 6.4~keV. Modelling of the EPIC PN data
suggest that this line consists of two components of approximately
equal strength: a broad line from an inclined accretion disc and a
narrow, unresolved line. Soft X-ray reflection from the ionised disc
accretion disc model adopted here also contributes to, but does not
completely explain, the broad, soft X-ray excess observed in Mrk~359.
Only weak features, due to L-shell absorption in iron M ions,
the \hbox{Ne\,{\sc IX}} and \hbox{O\,{\sc vii}} triplets in emission and
\hbox{O\,{\sc viii}} Ly$\alpha$ emission are seen superimposed on the
soft X-ray excess. Such soft X-ray excesses have now been clearly
detected in a number of broad-line AGN observed by {\it XMM-Newton},
and may well be ubiquitous (e.g. O'Brien et al. 2001; Reeves et al.
2001; Pounds et al. 2001). It is tempting to ascribe this emission as
the high-energy tail of the `big blue bump' powered by thermal
emission from the accretion disc.

A narrow 6.4~keV emission-line seems to be similarly ubiquitous in
broad-line AGN (Reeves et al. 2001; Yaqoob et al. 2001; Kaspi et al.
2001; Pounds et al. 2001; Lubinski \& Zdziarski 2001). The strength of
this line in a fairly low-luminosity AGN like Mrk~359 (with an EW
$\approx 120$~eV) supports the trend, suggested from previous {\it
XMM-Newton} and {\it Chandra} observations, that the narrow-line is
relatively stronger in low compared to high-luminosity AGN. Compared
to Mrk~359, NGC~5548 (Yaqoob et al. 2001) and NGC~3783 (Kaspi et al.
2001), the narrow-line is weaker in the luminous Seyfert~1 galaxies,
Mrk~205 (Reeves et al. 2001) and Mrk~509 (Pounds et al. 2001), while
it is not detected (EW $< 10$~eV) in the high-luminosity QSOs
PKS~0558$-$504 (O'Brien et al. 2001), S5~0836+710 and PKS~2149$-$306
(Fang et al. 2001). The narrow-line strength in the lower luminosity
AGN implies a significant solid angle (of order $\pi$ steradian) of
cool reprocessing gas lying outside our direct line of sight to the
X-ray source. While some of this may be BLR gas (e.g., Yaqoob et al.
2001), the most likely location is the putative molecular torus
surrounding the `central engine' (e.g. Ghisellini et al. 1994). 

A straightforward explanation for the correlation of narrow-line
strength and luminosity is then that the solid angle subtended by the
reprocessing gas declines in high-luminosity AGN, rendering the
reflected line difficult to detect. Such a trend would have
significant implications for the origin of the reprocessing gas, the
relative number of type-1 and type-2 AGN as a function of luminosity
and for models seeking to explain the spectral shape of the X-ray
background in terms of obscured AGN. For torus models in which the
reprocessing gas is a dusty, hydromagnetic wind, the reduction in
solid angle may be due to the field lines being `flattened' by
radiation pressure in high luminosity AGN (Konigl \& Kartje 1994).
Alternatively, Ohsuga \& Umemura (2001) suggest that dusty gas-walls
supported by radiation pressure from a circumnuclear starburst are
more common in relatively low luminosity AGN. Radiation pressure may
prevent such walls forming in high luminosity AGN.

Overall, the unusual optical spectrum of Mrk~359 does not translate to
an unusual X-ray spectrum. Rather, Mrk~359 has an X-ray continuum and
emission-line spectrum consistent with that of a low-luminosity
Seyfert 1 galaxy viewed fairly face-on. X-ray spectra for a larger
sample of the narrowest-lined AGN are required to determine whether
they have systematically different X-ray properties, such as more
face-on discs than typical Seyfert 1s or lower/higher accretion rates
than for other NLS1s.

\section*{Acknowledgements}

This work is based on observations obtained with the XMM-Newton, an
ESA science mission with instruments and contributions diectly funded
by ESA member states and the USA (NASA). The authors thank Sergei
Nayakshin for providing Xion and for helpful advice on accretion disc
models. JR is supported by a Leverhulme Research Fellowship and KP by
a Research Studentship from the UK Particle Physics and Astronomy
Research Council.

\end{document}